# Exchange Bias and Vertical Shift in $CoFe_2O_4$ nanoparticles

A. Mumtaz, K. Maaz, B. Janjua and S.K. Hasanain.

*Department of Physics, Quaid-i-Azam University, Islamabad, Pakistan*

## Abstract

Magnetic properties of core-shell cobalt ferrite nanoparticles 15 to 48nm prepared by a sol-gel route have been studied. It is shown that the coercivity follows non-monotonic size dependence varying as *1/d* above the maximum (*d* is the particle size). Field cooled magnetization exhibited both horizontal (exchange bias) and vertical shifts. The exchange bias is understood as originating at the interface between a surface region with structural and spin disorder and a core ferrimagnetic region. The dependence of the exchange bias and vertical shifts on the particle sizes and cooling fields are found to have significant differences and the differences are explained in the light of recent results which suggest that both weakly and strongly pinned spins are present at the interface. It is suggested that the exchange bias is dominated by the weakly pinned spins while the vertical shift is affected by the strongly pinned ones.

Corresponding author: arif@qau.edu.pk (A. Mumtaz)

# Introduction

Metal-oxide nanoparticles are currently a subject of immense interest because of their unusual optical, electronic and magnetic properties, which often differ from their bulk counterparts. Many of these unique properties also make them very promising candidates for a variety of applications in biomedical as well recording technology. Cobalt ferrite is an ideal candidate for many applications. While having large magnetocrystalline anisotropy and a wide size range for single domain behavior, it shows superparamagnetic behavior for sufficiently small particles at room temperatures. These features are relevant for magnetic recording and biomedical applications respectively.

However, even the early reports have shown the magnetic behavior of nanoparticles to be generally very complicated e.g., Berkowitz et al [1] have reported that the saturation magnetization of nanoparticles reduces as the particle size decreases. To explain this trend a simple model considering spin canting at the surface of the particles was proposed [2]. Recently polarized neutron experiment by Lin et al. [3] have demonstrated that a layer having thickness of about 1.2 nm, of magnetically disordered surface spin exists in $CoFe_2O_4$ nanoparticles. Kodama et al. have discussed number of reasons that result in surface spin disorder. In case of ionic compounds, the superexchange interaction between magnetic cations is antiferromagnetic, however, the ferrimagnetic order arises as the intersublattice exchange is stronger than the intrasublattice exchange. Kodama et al [4] have argued that the variation in the surface cations results in a distribution of net exchange fields, both positive and negative with respect to a cation's sublattices. Since the interaction is mediated by oxygen, and if some oxygen ions are missing from the surface, exchange bond will break and reduce the effective coordination of the surface ions. Similar results are expected if the surface is coated by organic molecules, since the electrons involved may not participate in superexchange. Using these arguments, Kodama et al. have proposed a model of the magnetization within these particles consisting of ferrimagnetically aligned core and a spin glass like surface layer. Such a physical picture is usually referred to in the literature as a core shell model with the unique features that the core and the outer shell may be magnetically different e.g ferromagnetic and ferrimagnetic respectively. It has been shown that such a two phase nanoparticle system mimics the interface effects as in the case of FM and AFM bilayer systems and may exhibit technologically and scientifically important effects such as horizontal and vertical loop shifts (exchange bias and vertical magnetization shift). How exactly these effects take place and their dependence on particle size and morphology etc are yet to be clearly understood and controlled.

In this work we report effects of the size on the magnetic properties of cobalt ferrite nanoparticles. This includes the variation of coercivity ($H_c$), saturation moment ($M_s$) and exchange bias as the average size of the nanoparticle is reduced from 48nm to 15 nm. Furthermore the exchange bias and vertical shift effects are discussed with respect to the role of the interface pinning strengths and their role in determining the size and field dependence of these features.

## Results and Discussion

The samples reported in this study were synthesized by sol gel technique described elsewhere [5]. Starting materials were Ferric chloride ($FeCl_2$), cobalt chloride ($CoCl_2$) and sodium dodecyle sulphate (NaDS) was selected as a surfactant and double distilled, de-ionized water was used as a solvent. During synthesis of all particles pH was monitored and controlled at different level for different particle sizes. The reaction was carried out at 80$^{o}$C. After the completion of reaction the solution was cooled and centrifuged at 3000 rpm two or three times in order to isolate the supernatants. The final product was grinded and annealed at various temperatures to obtain different particle sizes.

X-ray diffraction of the samples was obtained and $CoFe_2O_4$ phase was confirmed, no other impurities were detected. The size of the particles is determined by using Scherrer formula. The grain size obtained from this method was also confirmed using TEM images and we find that the grain size correlates well with that obtained from TEM analysis. Figure 1, shows a TEM images of one of the sample annealed at 600$^{o}$C for 10 hours with average particle size of 21 nm, which is in good agreement with XRD size calculations. Some moderately agglomerated particles as well as separated particles are also present in the images. The size control was made either by controlling the pH of the solution or by annealing temperature. We find that the size increases with increasing annealing temperature and the dependence is linear.

Magnetic measurements were made on vibrating sample magnetometer (VSM) at room temperature and 77 K with maximum applied field of 17 kOe. The hysteresis loops for one of the representative samples (d = 21 nm) at 77K and 300K are plotted in figure 2. It is observed that the hysteresis increases substantially at low temperature. It was found that full true saturation was not achieved in our samples even at 17 kOe, therefore the magnetization at the highest field (i.e.17 kOe) is referred to as the "saturation magnetization". The saturation magnetization ($M_s$ at 17kOe) as found at room temperature was 68 emu/g and remanent magnetization was 33 emu/g while at 77 K these values are 52 emu/g and 44 emu/g respectively. The saturation magnetization at 77K is significantly lower as compared to its room temperature value evidently due to the inability to rotate the moments at low temperatures. In the Stoner-Wohlfarth picture this inability to saturate the moments is attributed to the largeness of the ratio of effective energy barrier (Anisotropy energy –Zeeman energy) to the thermal energy KT. This does not exclude the possibility that the anisotropy energy may itself increase substantially at low temperatures. In Fig. 3 we display the dependence of saturation magnetization on particle size and $M_s$ is observed to decrease consistently with decreasing particle size. This effect can be related to the development, in ferrite nanoparticles, of a disordered or spin glass type layer at the surface [1, 4, 6]. For smaller particles the effects of this surface layer of reduced magnetization would be more significant due to the larger fraction of total spins lying within the surface region and hence the total magnetization (ferrimagnetic core and surface contributions) may be expected to decrease at smaller particle sizes.

The coercive fields for different size particles are plotted at room temperature (300K) and 77K in Fig. 4(a,b). It is evident from the figures that for a given particle size the coercivity typically increases by almost a factor of ten between these temperatures. It is also evident that the coercivity displays nonmonotonic size dependence. With decreasing particle size the coercivity goes through a maximum, peaking at around 27nm. At 77K however the fractional decrease in coercivities for the smaller size particles appears to be much smaller than at 300K.

The behavior of $H_c$ can be understood with reference to the role of the surface anisotropy. It has been shown by B$\phi$dker et al. [7] that the effective anisotropy constant of nanoparticle increases with decreasing size. The phenomenological expression for the effective anisotropy of spherical particles may be written as $K_{eff} = K_v + (6/d) K_s$. Several other works including simulation and experimental have also supported this expression [8]. If we assume similar behavior in these nanoparticles i.e. an increase in the effective anisotropy energy with reducing particle size this would tend to increase the coercivity of the nanoparticles within the Stoner-Wohlfarth picture, $H_c=2K/M_s$, consistent with the behavior above the peak in Fig. 4. This increase will however not continue indefinitely and as the particle size decreases to a small enough value, thermal effects will take over. For particles below this critical size (say~26nm) the thermal energy becomes sufficient to overcome the anisotropy energy enabling the easier reversal of moments and leading to lower critical fields for these small sizes [9].

This fact is also supported by observing the particle size dependence of the blocking temperature. The blocking temperature $T_B$ is defined as the temperature where the zero field cooled moment of the particles exhibits a peak versus temperature, for the reasons explained above. Typical ZFC M(T) behavior in a field of 5kOe is shown in the inset of Fig.5 while the main figure shows the variation of the peak temperature $T_B$ for different size particles. It is clear from fig.5 that blocking temperature decreases sharply for samples with size below ~26 nm, indicating that the thermal energy $K_BT$ remains sufficient to unblock the magnetic moment of these smaller particles down to lower temperature.

## Exchange Bias Effect

In the exchange bias studies the samples are cooled from room temperature ($T<T_c$) to a low temperature ($T<T_N$) in the presence of a magnetic field and the hysterias loops are recorded. The consequent shifts in M(H) loop have been extensively studied in the bi layer systems where an antiferromagnetic (strong anisotropy) material is deposited on top of a ferromagnetic layer and the horizontal shift of the loops has been termed as *exchange bias field* $H_{eb} = (H_{c2} - H_{c1})/2$ where $H_{c1}$ and $H_{c2}$ are the absolute values of positive and the negative coercive fields. A simple model explaining the exchange bias effect was proposed by Mieckeljohn and Bean [10]. In a bilayer type system of ferromagnetic (FM) and antiferromagnetic (AFM) layers, when the field is applied at temperature which is less than the curie temperature of the ferromagnetic layer and is above the Nèel temperature of the AFM layer, the spins of the low anisotropy FM material align with the applied field. When the temperature is lowered through the Neèl temperature of the AFM layer its spins align with respect to each other and may also couple with the FM spins depending upon the interfacial exchange coupling between the two layers. This will generate a uniaxial anisotropy parallel to the cooling field direction. This is due to the large anisotropy of the AFM layer that prevents the AFM spin rotation and these in turn prevent the FM spins in turning away from the cooling direction. Furthermore, it has been shown [11,12] that the hysteresis loop may also show a vertical shift of the magnetization that is related to the presence of pinned interfacial spins. A positive (upward) shift is attributed to a ferromagnetic interface coupling and a negative shift for an antiferromagnetic interface coupling. [12].

We have made a systematic study of the exchange bias effects in our system as a function of particle size, cooling field and number of field cycling. (The cycle number refers to the successive field cycle to which the sample was exposed after field cooling). In Fig. 6 the data is shown for a sample with average grain size of 21 nm, cooled to 77 K in a field of 12 kOe. In this figure we observe that the magnetization loop is shifted or displaced *both horizontally and vertically.* The horizontal shift or the exchange bias in figure 6 for a 21 nm particles is 750 Oe, when cooled from room temperature to 77 K in a field of 12 kOe.

The exchange bias for different size particles (15 – 48 nm) obtained as described above is plotted in figure 7. It is evident from the figure that the exchange bias has a nonmonotonic size dependence, similar to what is observed in coercivity field for these particles. The exchange bias first increases with increasing particle size, goes through a peak, decreases and eventually assumes negligible value for sizes in excess of 45nm. The peak position in both cases viz. $H_c$ and $H_{eb}$ is around 26-30 nm, suggesting a similar underlying mechanism.

Analyzing the behavior along the general lines of the MB model we anticipate that exchange bias in cobalt ferrite nanoparticles arises due to the interaction between the core and surface spins [13]. However in this case the core spins are aligned ferrimagnetically while at the surface the situation may be quite complex. In general the atomic coordination number of the surface spins is different from that of the core and this variation of atomic coordination number causes perturbations in the crystal field which destabilizes the ferrimagnetic order (at the surface). The spin alignments can take a multiplicity of forms [14] with several different ground states and resultantly the surface layer acts in a "spin glass like" manner [15]. Therefore, the particles behave as core-shell system similar to the metal-metal oxide nanoparticles with an oxide shell covering the metallic core [16]. The core shell interaction at the interface gives rise to the phenomenon of exchange bias effect.

The observed size dependence of the exchange bias can be explained on the basis of the typical models. The exchange bias $H_E$ has been estimated [15] to vary as

$$H_E = \frac{\sigma_{\mathrm{int}}}{M_{FM}\, t_{FM}},$$

where $\sigma_{int}$ is the interfacial exchange energy, $t_{FM}$ is the thickness of the FM core and $M_{FM}$ its magnetization. The 1/d dependence of the exchange bias in the size region above the peak in $H_E$ is consistent with the thickness of the increase of the FM core with increasing particle size. The decrease of the exchange bias below the maximum can be explained in the light of the same argument that explains the variation of the coercivity. The exchange bias gets destabilized in the smaller particles due to the thermal fluctuations of the core and shell spins. Another explanation is offered by the model of [17]. In this picture the condition for the observation of the exchange bias is that $E_{\mathrm{int}}<E_Z$, where $E_Z$ is the effective Zeeman energy $E_Z=MHV$ while $E_{\mathrm{int}}$ is the interface exchange energy $E_{\mathrm{int}}$. It is possible that the decrease of the exchange bias at smaller sizes is due to the smallness of the volume dependant Zeeman energy as compared to the anisotropy energy which is expected to be large for the smaller size particles due to the 1/d dependence of the surface energy mentioned earlier. In general both effects may contribute to the weakening of the exchange bias for very small particles.

# Vertical Shift in Magnetization

The presence of a vertical shift accompanying the exchange bias has been associated with the presence of uncompensated pinned spins at the interface of the two magnetic media [18, 19]. In our nanoparticles we have observed a positive vertical shift that indicates a positive interaction between the two types of spins across the interface. A systematic trend is seen for the variation of the vertical shift δM with the size of the nanoparticles. We refer to figure 8, for the size dependence of this shift. It is apparent that it shows a peak coinciding with the peak in $H_{ex}$. Furthermore δM is very small for larger particles e.g., for particles of size 32nm and larger it is less than 10 emu/gm, whereas a very sharp jump is observed between 27 and 32nm where a peak (40emu/gm) is observed. Below this peak the vertical shift again decreases strongly for smaller size particles. It is noticeable that despite the similarity of the trends in $H_{ex}$ and δM the drop in the vertical shift with increasing size beyond the peak is very rapid and δM becomes almost negligible for the sizes where the exchange bias is still very significant. Thus it appears that the size increase beyond the maximum has a very drastic effect on the VS and to a smaller extent on the exchange bias. The two however appear to vanish at the same size of about 42nm. The differing trends between the exchange bias and vertical shift have also been observed by one of us in the Fe particles with a core-shell structure [20]. For larger sized Fe particles (14nm) the exchange bias was observed to remain nonzero while the vertical shift was completely absent. Smaller sized particles (9nm) on the other hand were seen to exhibit both effects.

The exchange bias and vertical shifts were also studied as a function of the number of cooling field cycles and cooling field magnitude. The effect of successive field cycles was to decrease both the exchange bias and vertical shift in a similar manner. The effect of the cooling field magnitude is shown in fig. 9. We note that while both continue to rise for low fields (H<5kOe), the vertical shift becomes almost constant while the exchange bias declines very significantly above this field. The initial rise with field is understood to occur as the field initially is not sufficient to align all the ferrimagnetic spins. With increasing fields more of the core spins effectively participate in the exchange bias. Similarly larger cooling fields presumably enable more of the interfacial spins to become decoupled from the antiferromagnetic shell and be pinned with the interface at directions close to the field direction thereby increasing the vertical shift. At a large enough magnitude of the cooling field the reversible magnetization, that includes the core spins and a fraction of the weakly pinned spins of the spin glass like shell, increases [16] thereby reducing $H_{EB}$ according to the relation $H_E = \frac{\sigma_{\text{int}}}{M_{FM}\, t_{FM}}$ where the reversible magnetization plays the role of $M_{FM}$.

The relative insensitivity of the vertical shift to higher fields in our data suggests that the spins responsible for it could remain unchanged until such fields that the Zeeman energy becomes comparable to the pinning energy of these spins at the interface. More intriguing is the trend of the mismatch in the behavior of $\delta M$ and $H_{ex}$ which appears to be similar to the trend between the remanent moment and exchange bias respectively as functions of the cooling field reported by [21]. We note that the differing trends in the two viz. $\delta M$ and $H_{ex}$ with cooling fields and particle sizes is inconsistent [22] with a picture such as the M-B model where all the spins at the interface have a similar pinning or with models where the same type of spins (uncompensated strongly pinned) are responsible for both $\delta M$ and $H_{ex}$ [18,19]. It appears that other considerations regarding the spins involved in the two effects also have to be taken into account. We refer to the discussion in [22] who based on their observations on soft X-ray resonant magnetic scattering studies on Fe/CoO exchange bias system. They find strong evidence for *both* strongly and weakly pinned pins at the AFM/FM interface. They argue that structural defects, etc. cause non-ideal magnetic interfaces with statistically a fraction of the AF spins having lower anisotropy as compared to the bulk AFM ones. These weaker pinning interfacial AF spins can rotate together with the ferromagnet and lead to the exchange coupling and induce an enhanced coercivity. Alongside, as assumed in the original Meikeljohn and Bean model, there are also the uncompensated spins at the interface with large anisotropy constants ($\sim K_{AF}$). A similar model can account for our observations where the vertical shift may be attributed to the large anisotropy (strongly pinned) spins while the exchange bias is to be attributed to spins with reduced anisotropy constants and pinning. We suggest that with increasing particle size the fraction of the strongly pinned interfacial spins decreases faster than the total uncompensated spins at the interface leading to a more rapid decline in the vertical shift than the exchange bias. Similarly, large fields are able to rotate the weaker pinned spins at the interface and hence decrease the exchange bias while leaving the strongly pinned spins responsible for the vertical shift relatively unaffected.

# Conclusion

Various magnetic properties of cobalt ferrite nanoparticles have been studied and explained with respect to variations in particle sizes and the role of bulk and surface anisotropy. The coercivity exhibits a 1/d variation that is explained in the light of the 1/d dependence of the surface anisotropy. The exchange bias in the system is understood as originating in the structurally disordered surface region with multiple spin configurations leading to a spin glass like magnetic response. The magnetic exchange between the surface spin glass like spins and the core ferrimagnetic spins is responsible for the horizontal loop shifting. Vertical shifts in the field cooled magnetization have also been studied and the most significant part of our work is the difference in the trends of the exchange bias and vertical shifts as functions of particle size and cooling fields. These differences have been explained as originating in the two types of spins at the interface viz weakly pinned spins that appear to be responsible for the exchange bias and the strongly pinned ones principally responsible for the vertical shift.

## Acknowledgment

K.M and A. M wish to acknowledge the Higher Education Commission (HEC) for providing PhD fellowship and Research Grant (*No. 20-74/Acad(R)/03*) for enabling this work. S.K.H also acknowledges the H.E.C for Research Grant (*No. 20-80/Acad(R)/03)* on ferromagnetic nanomaterials.

## References

[1] A.E. Berkowitz, J.A. Lahut, I.S. Jacobs, L.M. Levinson, D.W. Forester, Phys. Rev. Lett. **34**, 594 (1975) and A.E. Berkowitz, J.A. Lahut, C.E. Van Buren, IEEE Trans. Magn. MAG-**16**, 184 (1980).

[2] J.M.D. Coey, Phys. Rev. Lett. **27**, 1140 (1971).

[3] D.Lin, A.C.Nunes, C.F.Majkrzak, A.E.Berkowitz, Journal of Magnetism and Magnetic Materials, **145**, 343 (1995).

[4] R. H. Kodama, A. E. Berkowitz, Phys. Rev. Lett., **77**, 394(1996).

[5] K.Maaz, A.Mumtaz, S.K.Hasanain, A.Ceylan, Journal of Magnetism and Magnetic Materials (submitted for publication).

[6] A.C. Nunes, J. Appl. Cryst.., **21**, 129 (1998).

[7] F. Bφdker, S. Mφrup, S. Linderoth, Phy. Rev. Lett., **72**, 282 (1994).

[8] B.R. Pujada, E.H.C.P. Sinnecker, A.M. Rossi, A.P. Guimaraes, J. App. Phy., **93**, 7217 (2003).

[9] H. Pfeiffer, Phy. Stat. Sol. (a) **118**, 295 (1990).

[10] W.H. Meiklejohn, C.P. Bean, Phys. Rev., **105**, 904 (1957) and W.H. Meiklejohn, C.P. Bean, Phys. Rev., **102**,1413 (1956) .

[11] M. Verelst, T.O. Ely, C. Amiens, E. Snoeck, P. Lecante, A. Mosset, M. Repaud, J.M. Broto and B. Chaudret, Chem. Mater. 11, 2702 (1999).

[12] J. Nogues, C. Leighton and I.K. Schuller, Phys. Rev. B, **61** 1315 (2000).

[13] R.H. Kodama, A.E. Berkowitz, J. Appl. Phys. **81**, 5552 (1997).

[14] R.H. Kodama, Journal of Magnetism and Magnetic Materials, **200**, 359 (1999).

[15] J. Nogues, Ivan K. Schuller, Journal of Magnetism and Magnetic Materials, **192**, 203 (1999).

[16] R.K. Zheng, G.H. Wen, K.K. Fung, X.X. Zhang, J. Appl. Phys. **95**, 5244 (2004).

[17] A.N. Dobrynin, D.N. Ievlen, K. Temst, P. Lievens, J. Margueritat, J. Gonzalo, and C.N. Afonso, S.Q. Zhou, A. Vantomme, E. Piscopiello and G. Van Tendeloo, Appl. Phys. Letters., **87**, 012501-1 (2005).

[18] H. Ohldag,, A. Scholl, F. Nolting, E. Arenholz, S. Maat, A.T. Young, M. Carey, and J. Stohr, Phys. Rev. Lett. **91**, 017203-1 (2003).

[19] Oscar Iglesias, Xavier Batlle and Amilcar Labarta, arXiv: Cond-Mat./0509553 **v2** (2005).

[20] A. Ceylan, C.C. Baker, S.K. Hasanain, S. Ismat Shah. Journal of Applied Physics (submitted for publication).

[21] L.D. Bianco, D. Fiorani, A.M. Testa, E. Bonetti, L. Signorini, Phys. Rev. B, **70**, 052401-1, (2004).

[22] Florin Radu, A. Nefedov, J. Grabis, G. Nowak, A. Germann, H. Zebal, Journal of Magnetism and Magnetic Materials, **300**, 206 (2006).

## Figures Captions

Fig. 1: TEM micrograph of $CoFe_2O_4$ nanoparticles annealed at $600^oC$ for 10 hrs with average crystallite size of 21 nm.

Fig. 2. Hysteresis loops for 21nm $CoFe_2O_4$ nanoparticles at room temperature (300K) and at 77K after zero field cooling (ZFC).

Fig. 3. Saturation magnetization ($M_S$) as a function of particle size. Ms is defined as the moment at 15 kOe.

Fig. 4a. Variation of the coercivity ($H_C$) with mean particle size at room temperature. The peak is evident. Inset shows the variation of $H_c$ with 1/d at room temperature for d > 21nm.

Fig. 4b. Variation of the coercivity ($H_C$) with mean particle size at 77K.

Fig. 5. Blocking temperature as a function of particle size. The inset shows a typical M(T) curve for 21nm particles after field cooling at 5kOe.

Fig. 6. Hysteresis loops for 21nm $CoFe_2O_4$ nanoparticles at 77K after field cooling in 12 kOe. The asymmetry of the loop is evident along both horizontal and vertical axes.

Fig.7. The variation of exchange bias ($H_{ex}$) with mean particle diameter (nm).

Fig. 8. The dependence of exchange field ($H_{ex}$) and vertical shift (δM) on particle size. The very rapid decline of δM above the peak is to be compared to the slow decrease of $H_{ex}$. (See text for details)

Fig. 9. The dependence of exchange field ($H_{ex}$) and vertical shift (δM) on the cooling field. The relative intensity of δM and the decline of $H_{ex}$ at high field is evident

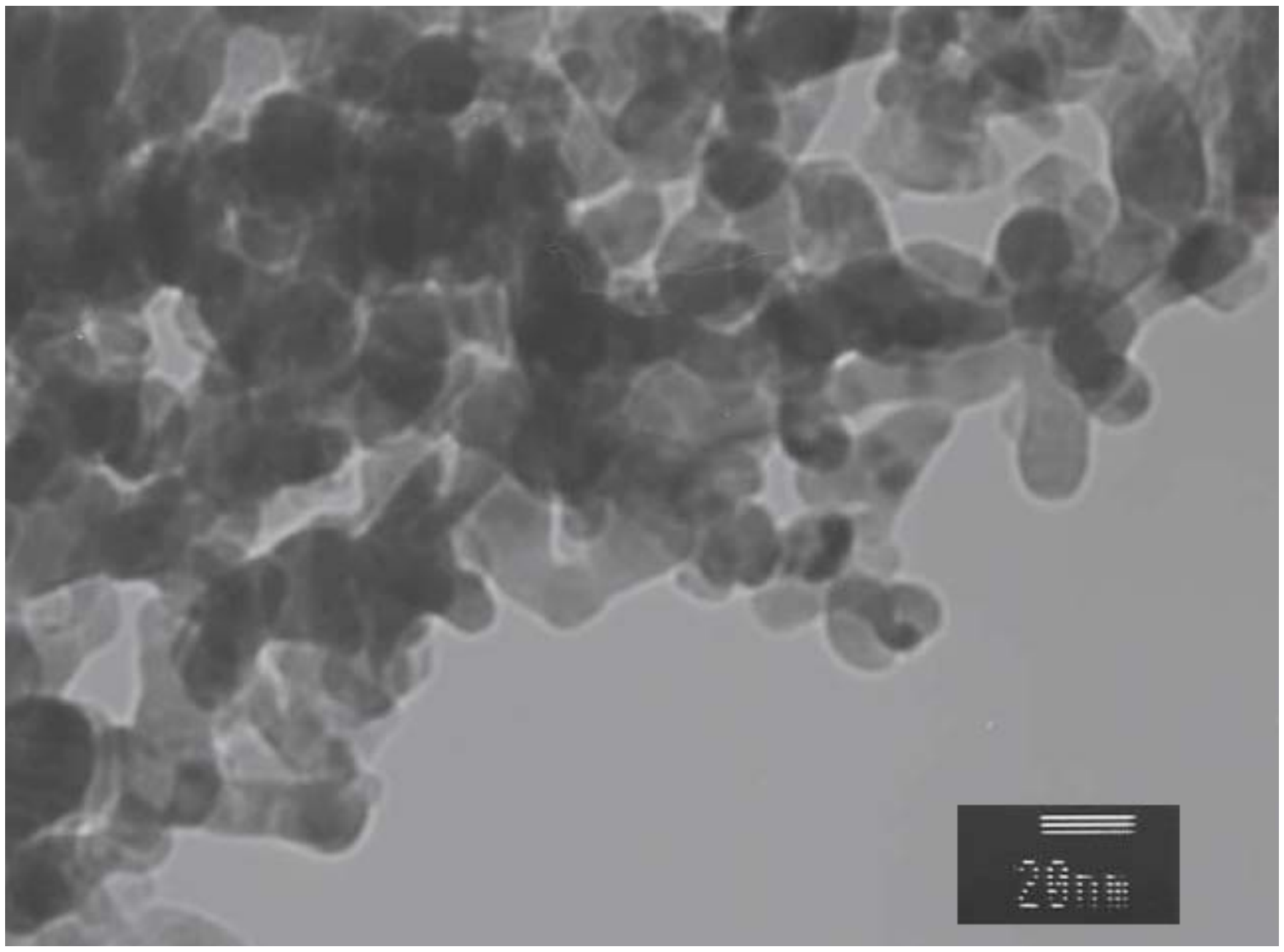

FIG.1. A. Mumtaz, et al. submitted to PRB

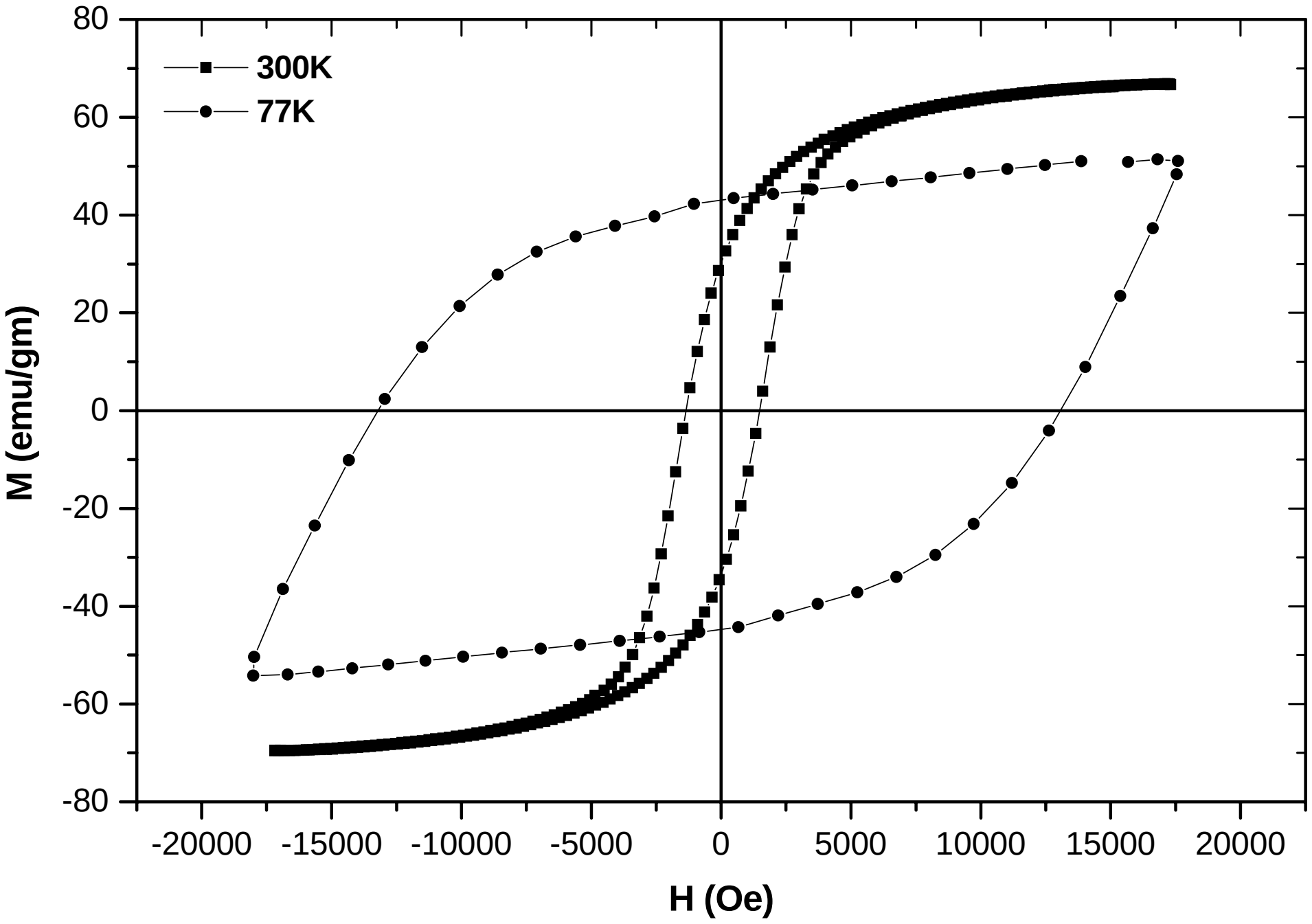

FIG.2. A. Mumtaz, et al. submitted to PRB

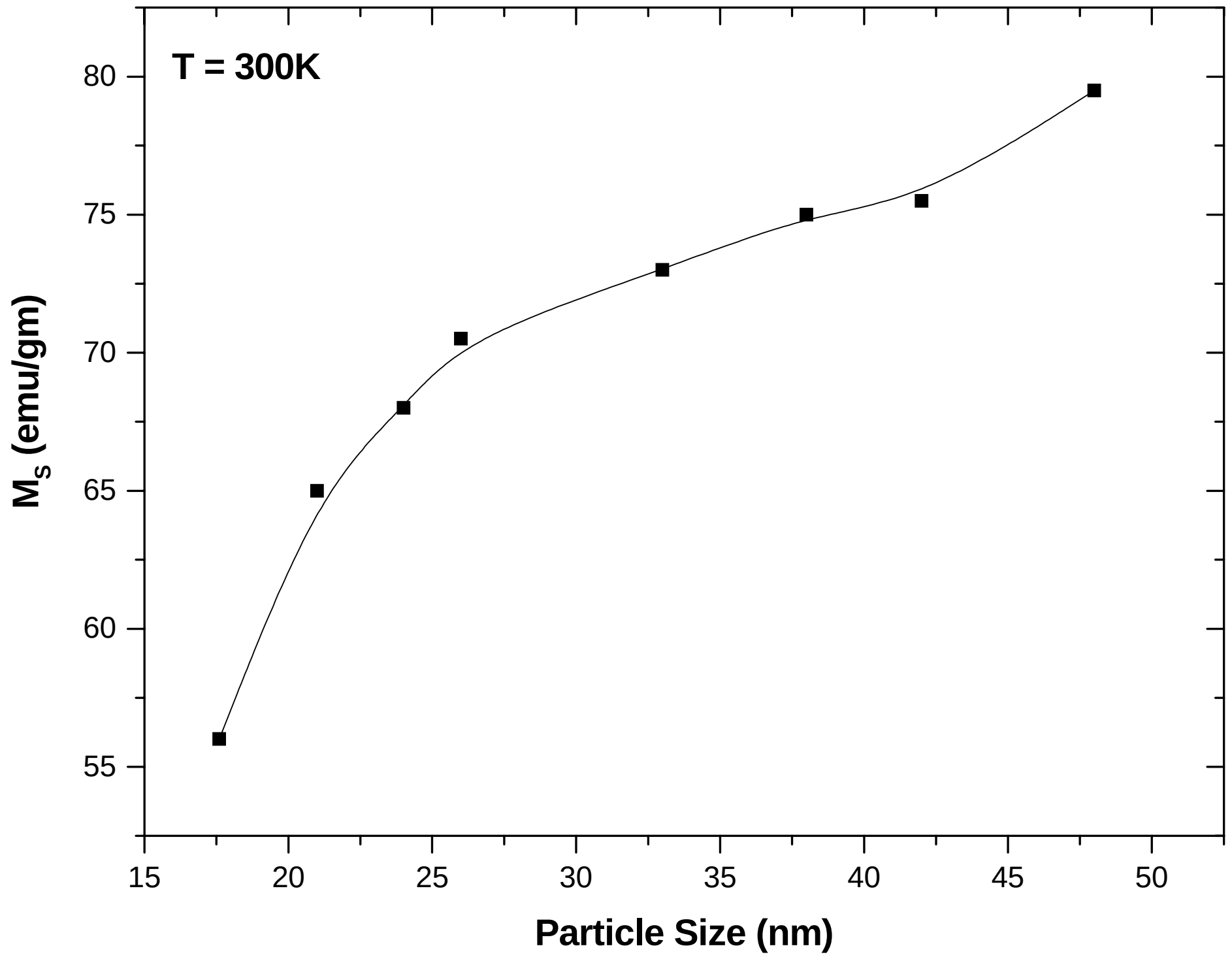

FIG.3. A. Mumtaz, et al. submitted to PRB

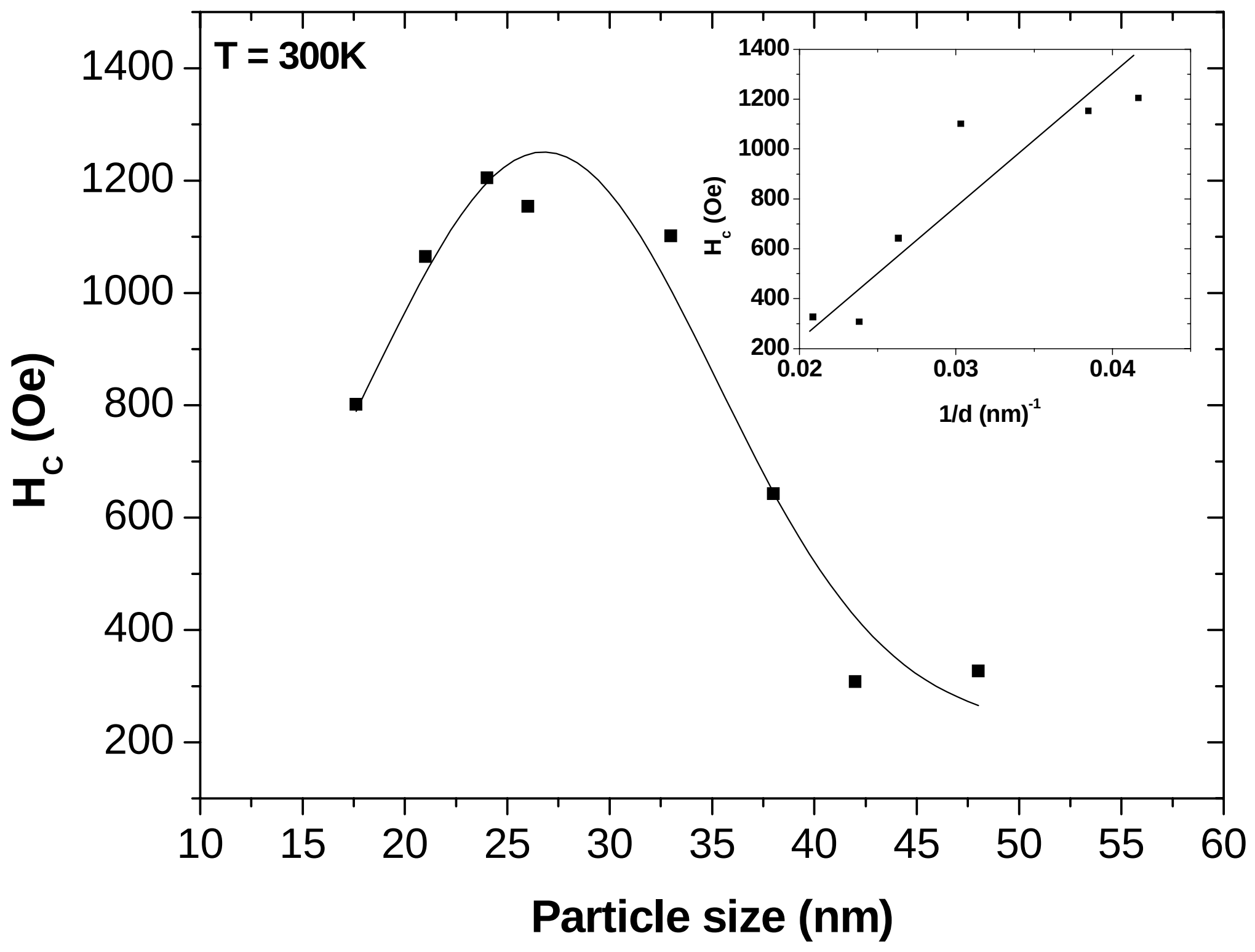

FIG.4a. A. Mumtaz, et al. submitted to PRB

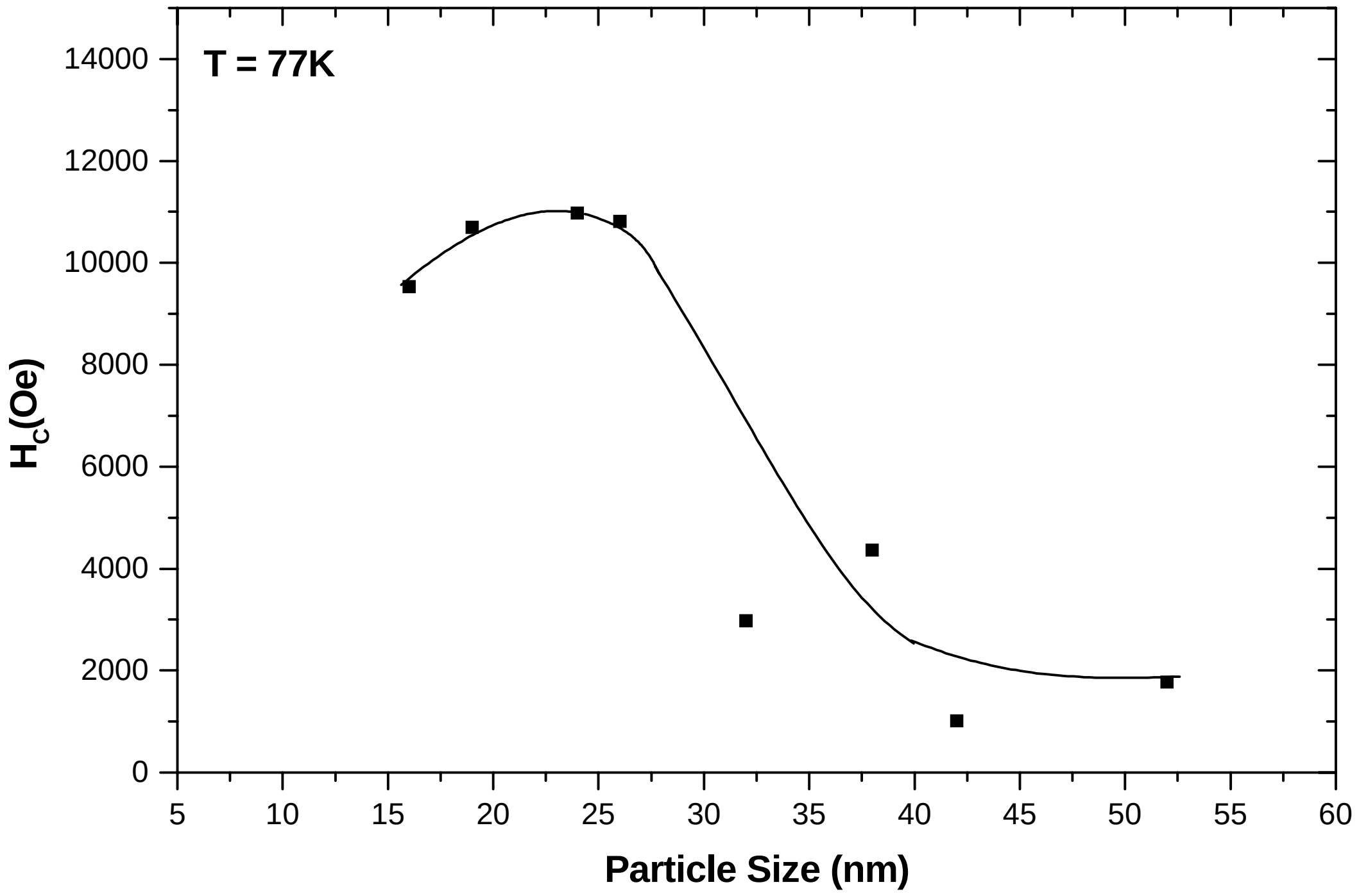

FIG.4b. A. Mumtaz, et al. submitted to PRB

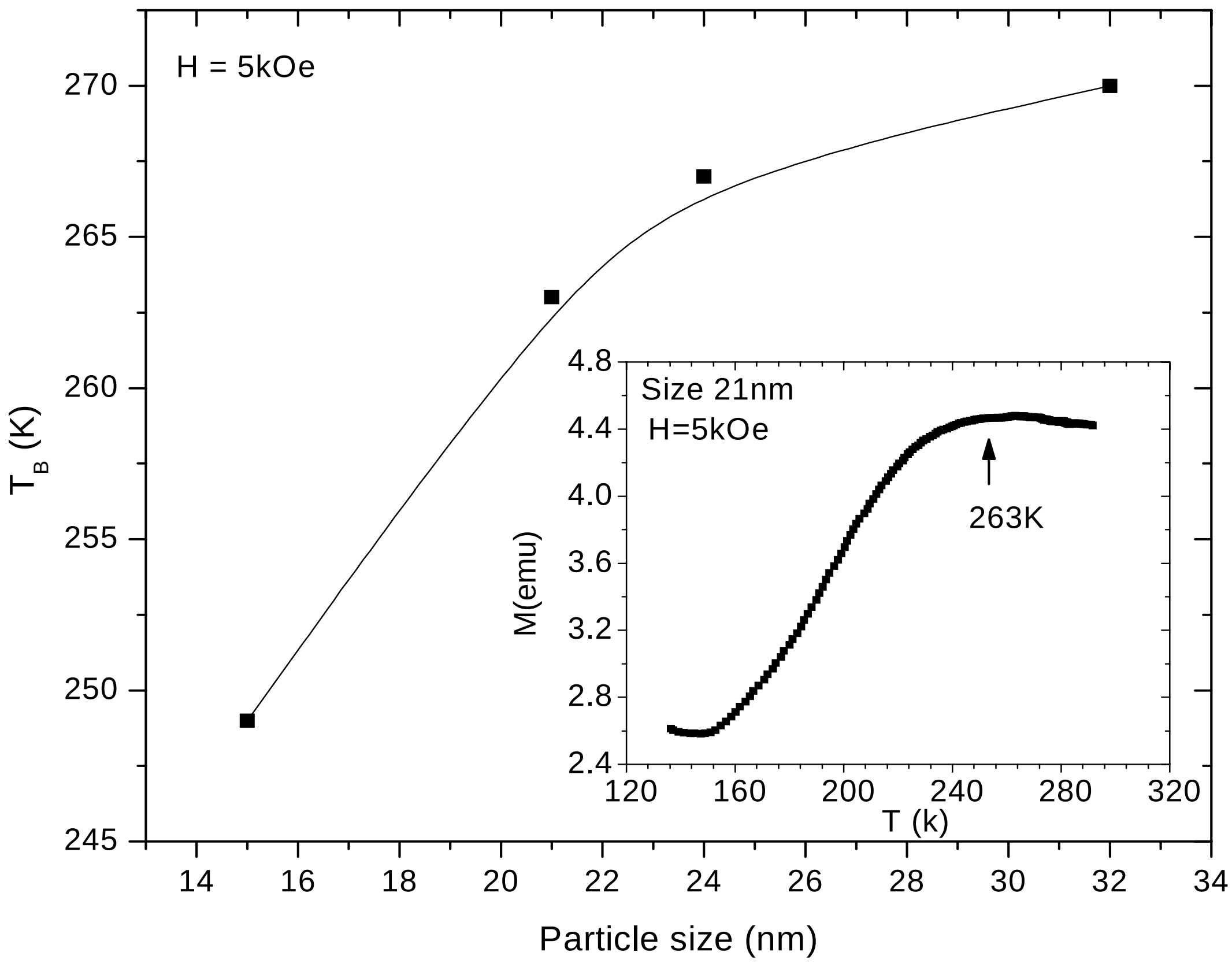

FIG.5. A. Mumtaz, et al. submitted to PRB

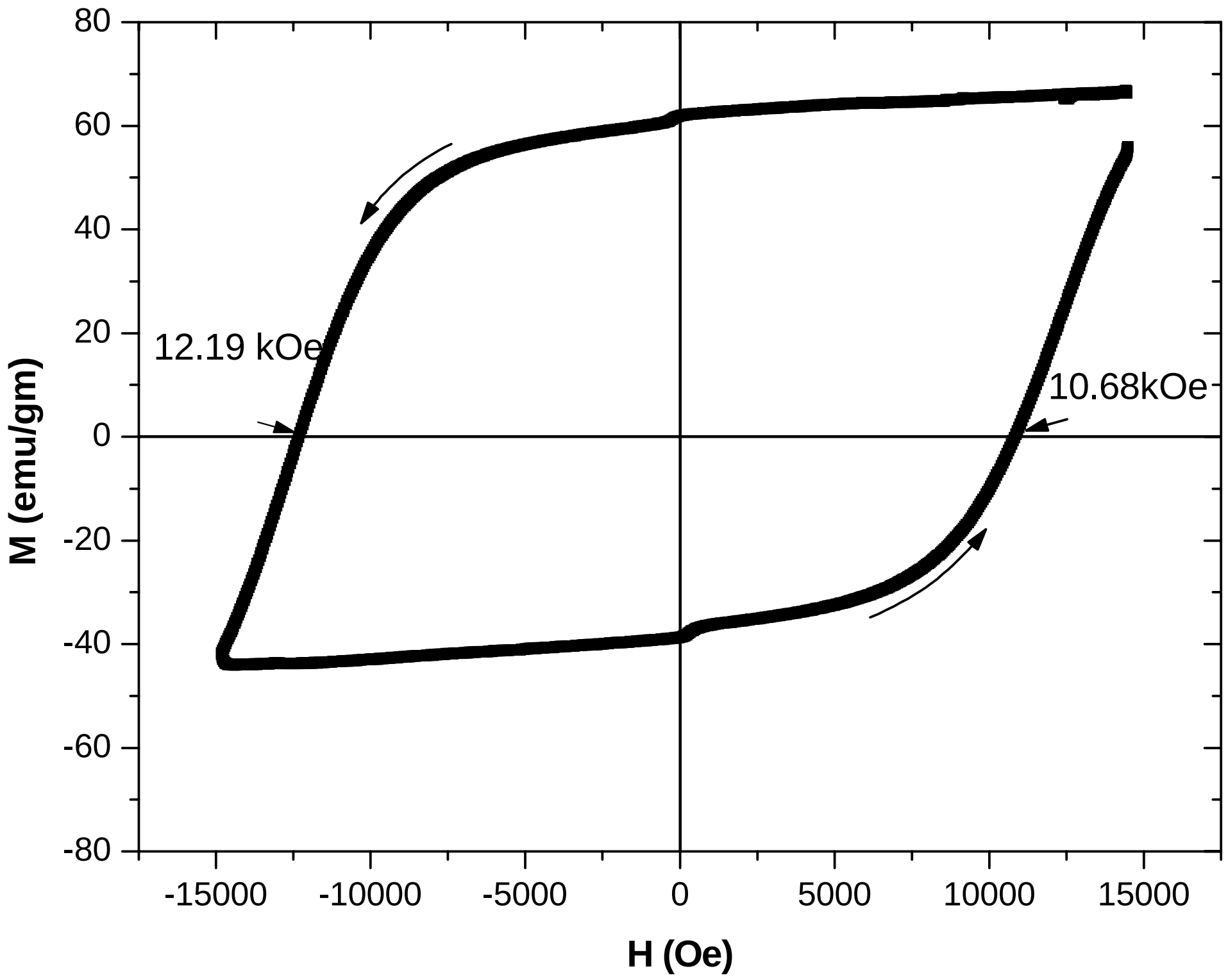

FIG.6. A. Mumtaz, et al. submitted to PRB

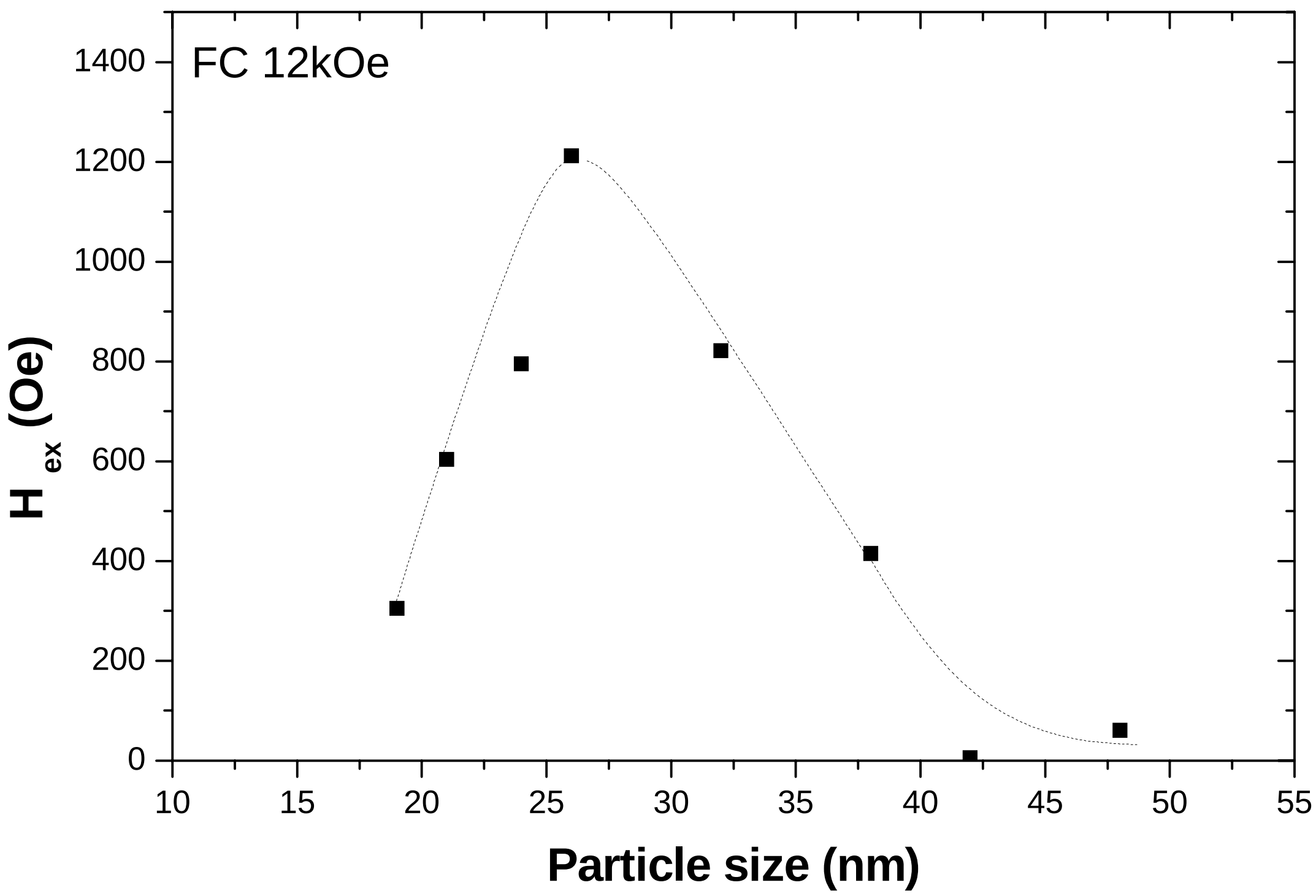

FIG.7. A. Mumtaz, et al. submitted to PRB

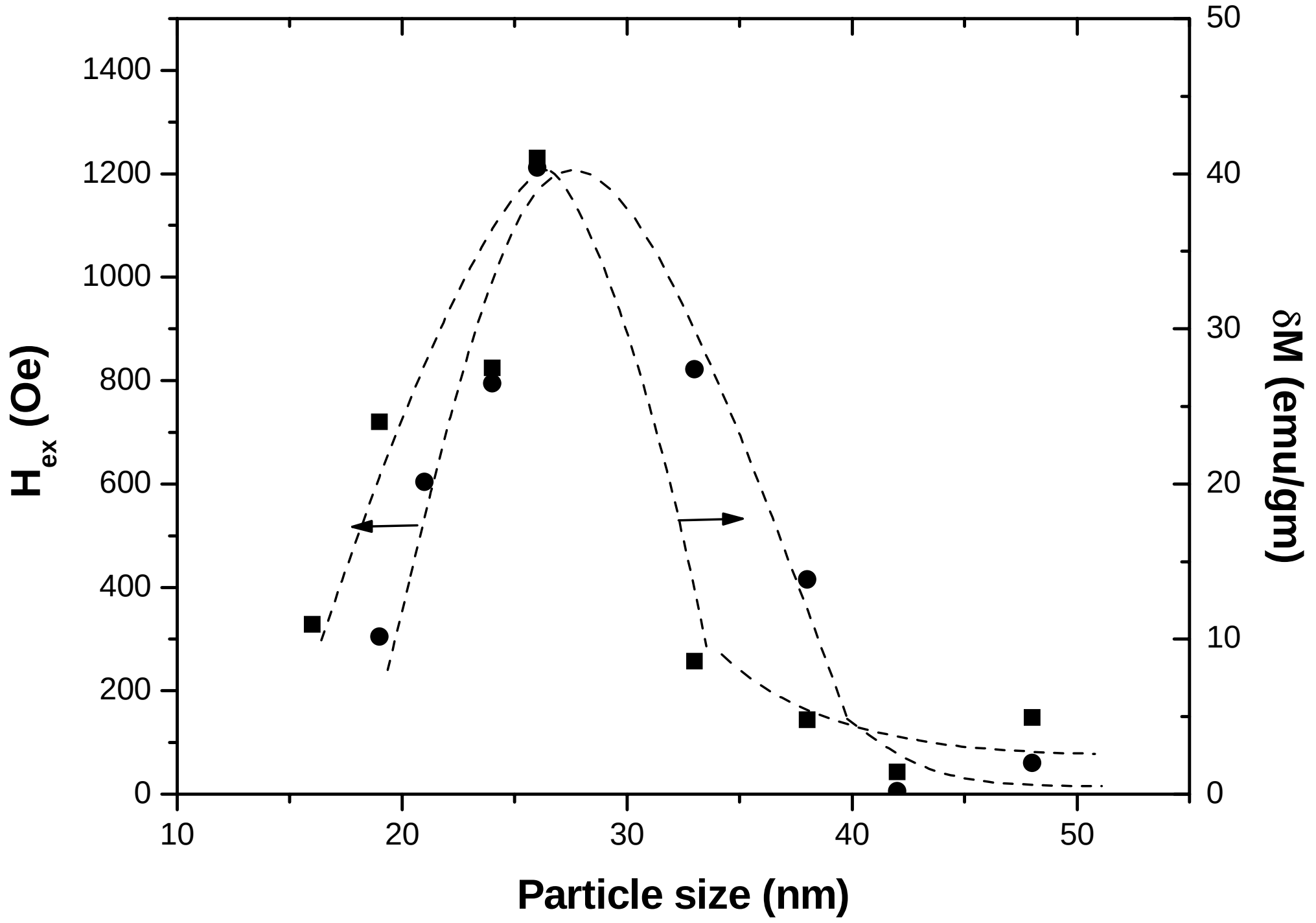

FIG.8. A. Mumtaz, et al. submitted to PRB

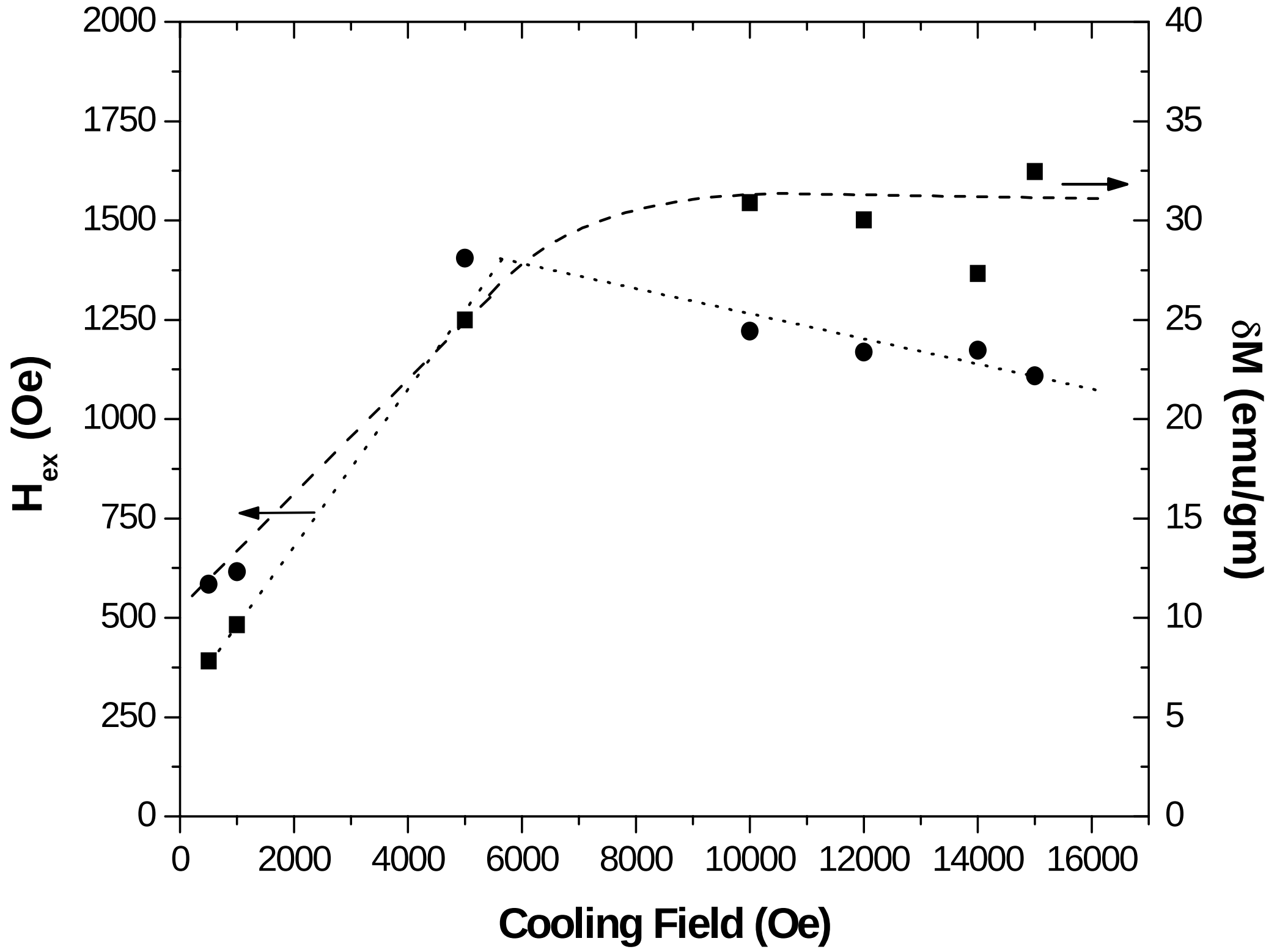

FIG.9. A. Mumtaz, et al. submitted to PRB